\documentclass{eptcs}
\usepackage[hide]{ed}
\usepackage[toc, page]{appendix}
\usepackage{amstext,amsmath,amssymb,doi,hyperref}
\usepackage{xspace,xcolor,realboxes}
\usepackage{listings}
\hypersetup{bookmarksopen,colorlinks,linkcolor=blue,citecolor=blue,bookmarksnumbered}

\usepackage{wrapfig,paralist,alltt,array,basics}

\usepackage{tikz}
\usetikzlibrary{mmt,fit,backgrounds,shapes,shadows,positioning}

\def\mmt{\textsc{Mmt}\xspace}
\def\omdoc{\textsf{OMDoc}\xspace}
\def\ommt{\omdoc/\mmt}
\def\mathhub{\textsf{MathHub}\xspace}

\definecolor{backcolor}{gray}{.92}

\newcommand{\expr}[1]{%
    \Colorbox{backcolor}{\ensuremath{\mathtt{ #1 }}}%
}

\title{Alignment-based Translations Across Formal Systems Using Interface Theories}
\author{Dennis M\"uller\institute{
  Computer Science, FAU Erlangen-N\"urnberg}
  \and
Colin Rothgang\institute{Mathematics, Jacobs University Bremen}
	\and
Yufei Liu\institute{Mathematics, Jacobs University Bremen}
	\and
Florian Rabe\institute{Computer Science, Jacobs University Bremen}
  }
  \def\titlerunning{Alignment-based Translations Across Formal Systems Using Interface Theories}
  \def\authorrunning{D. M\"uller, C. Rothgang, Y. Liu \& F. Rabe}

\begin{document}
\maketitle

\begin{abstract}
Translating expressions between different logics and theorem provers is notoriously and often prohibitively difficult, due to the large differences between the logical foundations, the implementations of the systems, and the structure of the respective libraries.
Practical solutions for exchanging theorems across theorem provers have remained both weak and brittle.
Consequently, libraries are not easily reusable across systems, and substantial effort must be spent on reformalizing and proving basic results in each system.
Notably, this problem exists already if we only try to exchange theorem \emph{statements} and forgo exchanging \emph{proofs}.

In previous work we introduced alignments as a lightweight standard for relating concepts across libraries and conjectured that it would provide a good base for translating expressions.
In this paper, we demonstrate the feasibility of this approach.
We use a foundationally uncommitted framework to write interface theories that abstract from logical foundation, implementation, and library structure.
Then we use alignments to record how the concepts in the interface theories are realized in several major proof assistant libraries, and we use that information to translate expressions across libraries.
Concretely, we present exemplary interface theories for several areas of mathematics and --- in total --- several hundred alignments that were found manually.
\end{abstract}

\section{Introduction and Related Work}\label{sec:intro}
 \paragraph{Motivation}
There is a vast plurality of formal systems and corresponding libraries for formalized knowledge.
However, almost all of these are non-interoperable because they are based on differing, mutually incompatible logical foundations (e.g. set theory, higher-order logic, variants of type theory etc.), implementations, and library structures, and much work is spent developing the respective basic libraries in each system.
Because a library in one such system is not reusable in another system, developers are forced to spend a large amount of time and energy developing parallel formalizations in multiple systems.

Ideally, given two or more libraries, one would want to integrate them with each other so that knowledge formalized in one of them can be reused in, translated to or identified with contents of the others.
In particular, translation across formal systems offers up many potential applications such as 
\begin{compactitem}
	\item accessing and using contents from other libraries during formalization,
	\item aiding premise selection by importing the premises used in proofs from other systems,
	\item presenting contents of an unfamiliar library using notations of a familiar one, or 
	\item freely combining user interfaces with libraries from different systems.
\end{compactitem}

\paragraph{Integrating Libraries}
There are two ways to take on library integration.
Firstly, we can try to translate the contents of one formal system directly into another system.
This requires intimate knowledge of both systems and their respective foundations used.

Secondly, we can use a \emph{logical framework} that provides a uniform intermediate data structure, in which we can specify the respective foundations and their libraries.
This approach has been used by the authors' research group in \cite{IanKohRab:tmmlof11} for Mizar, in \cite{KalRab:hollight:14} for HOL Light and in \cite{KohMueOwr:mpagsiuf17} for PVS, using the \mmt framework \cite{RabKoh:WSMSML13}.
A similar approach is currently underway using Dedukti \cite{dedukti} as the framework system.

Neither solution has proved particularly successful.
On the one hand, direct translations are expensive and must be developed for each pair of systems.
On the other hand, logical frameworks are a step in the right direction because they allow for a star-shaped integration architecture.
But they do not magically solve the library integration problem: logic and library translations remain extremely difficult.
A detailed analysis is given in \cite{KohRab:qrtpflmk15}.

In recent work we have come to understand this problem more clearly and suggested a systematic solution.
Firstly, in \cite{KohRabSac:fvip11} and \cite{KohRab:qrtpflmk15}, we developed the idea of \textbf{interface theories}.
Using an analogy to software engineering, we can think of interface theories as specifications and of theorem prover libraries as implementations of formal knowledge.
Libraries of interface theories must critically differ from typical theorem prover libraries: they must follow the axiomatic method (as opposed to the method of definitional extensions), be written with minimal foundational commitment, and not contain definitions or proofs.

Secondly, in \cite{KalKohMue:alfms16}, we introduced \textbf{alignments} as a primitive concept for lightweight translations between libraries.
In the simplest case, an alignment is a pair of symbol identifiers from two different libraries such that both symbols are ``morally the same'' mathematical concept.
In particular, two aligned symbols may use entirely different (possibly not even logically equivalent) definitions.
Specifically, alignments between an interface theory and a theorem prover provide the information how a library implements the concepts of the interface theory.

Thirdly, in the OpenDreamKit project~\cite{DehKohKon:iop16,OpenDreamKit:on} we pursue the same approach in the context of computer algebra systems.
In this context, we have already developed some interface theories for basic logical operations such as equality, with approximately 300 alignments to theorem prover libraries.

In this paper we follow up on this suggestion by designing several interface theories and finding their alignments to several major formal libraries.
Concretely, we have built theories for numbers, sets, lists, topology, combinatorics and analysis as representative examples of important domains.
Moreover, we have found alignments of the involved symbols with the libraries of HOL Light, PVS, and Mizar \ednote{Coq?}.

Consider for example the PVS symbol \expr{member} and the HOL Light symbol \expr{IN}.\ednote{mention mizar}
Both represent the mathematical concept of element-hood of (typed) sets, but they live in different libraries based on different foundations employed by different systems. 

We can see the two symbols being aligned via a sequence of consecutive steps of abstraction:
\begin{enumerate}
	\item We import the libraries of formal systems to a generic, foundationally neutral formal language (namely \ommt) with the help of a formalization of the system's primitives.
	\item We align the system specific symbols (which we collectively refer to as the \emph{system dialect})\label{lbl:dialect}
		with generic representations of the same concept. E.g. we align a symbol for HOL Light's function-application with a generic symbol for function application.
	\item Ultimately, we (possibly after additional steps) end up with the same purely mathematical concept
		expressed in a system- and foundationally neutral language.
\end{enumerate}

For the example with sets, we capture this with
\begin{compactitem}
 \item a symbol for elementhood in the interface theory for sets,
 \item two alignments of this new symbol with the symbols \expr{member} and \expr{IN}, respectively.
\end{compactitem}

This yields a star-shaped network as in the diagram on the right with various formal libraries on the outside, which are connected by alignments via representations of their system dialects (lower-case letters) to the interface theory in the center.

\ednote{use PVS, HOL Light, etc instead of A, B, ..., or drop the figure altogether}
\begin{wrapfigure}{r}{6.1cm}
\vspace{-1em}
 \begin{tikzpicture}[scale=1.5]
      \tikzstyle{withshadow}=[draw,drop shadow={opacity=.5},fill=white]
      \tikzstyle{system}=[draw]
      \tikzstyle{standard}=[circle,fill=blue!30]
      \tikzstyle{interface}=[circle,fill=purple!30,inner sep = 1pt,]
      \node[system] (a) at (-0.1,.3) {PVS};
      \node[system] (c) at (2.2,0.7) {HOL Light};
      \node[system] (e) at (1.1,2.7) {Mizar};
      \node[system] (g) at (-0.9,2) {Coq};
      \node[standard] (m) at (.5,1.5) {I};
      \node[interface] (ia) at (0.2,.9) {p};
      \node[interface] (ic) at (1.1,1.2) {h};
      \node[interface] (ie) at (.8,2.1) {m};
      \node[interface] (ig) at (-.1,1.75) {c};
      \draw (m) -- (ia) -- (a);
      \draw (m) -- (ic) -- (c);
      \draw (m) -- (ie) -- (e);
      \draw (m) -- (ig) -- (g);
      \begin{pgfonlayer}{background}
        \node[draw,cloud,fit=(ia) (ic) (ie) (ig),
                   inner sep=-7pt,withshadow] (st) {};
      \end{pgfonlayer}
\end{tikzpicture}
\end{wrapfigure}

\paragraph{Finding Alignments}
Even though we know that numerous alignments exist between libraries, not many alignments are known concretely or have been represented explicitly.
Therefore, a major initial investment is necessary to obtain a large library of interface theories and alignments.
There are three groups of alignment-finding approaches.

\emph{Human-based} approaches examine libraries and manually identify alignments.
This approach has been pursued ad hoc in various contexts.
For example, the library translation of \cite{Obua2006} included some alignments for HOL Light and Isa\-belle/HOL, which were later expanded by Kaliszyk.\ednote{citation?}
The Why3 and FoCaLiZe systems include alignments to various theorem provers that they use as backends.
\ednote{sledgehammer?}

The remaining two classes use \emph{artificial intelligence} methods.

\emph{Logical} approaches align two concepts if they satisfy the same theorems.
This cannot directly appeal to logical equivalence because that would require a translation between the corpora.
Instead, they compare the theorems that are explicitly stated in the corpora. 
The theorems should be normalized first to eliminate differences that do not affect logical equivalence.
The quality of the results depends on how completely those theorems characterize the respective concepts.
In well-curated libraries, this can be very high~\cite{KohMue:cicm15}.

\emph{Machine learning--based} approaches are inherently based on statistical patterns and hence naturally inexact.
The main research in this direction is carried out by Kaliszyk and others \cite{tgck-cicm14}.

While finding alignments automatically is promising for \emph{perfect alignments}, where two symbols only differ in name but are otherwise used identically, it is a different matter for \emph{imperfect} alignments.
For example, consider binary division which yields \textsf{undefined} or \textsf{0} when the divisor is zero, versus a strict division that requires an additional proof-argument that the divisor is nonzero. 
Here automation becomes much more difficult, because the imperfections often violate the conditions that an automatic approach uses to spot alignments.

We apply the human-based approach in this paper and demonstrate that it is feasible.
We present the largest set ever of systematically collected, human-verified alignments from multiple major proof assistants.
In addition to its inherent value, it will also greatly benefit the artificial intelligence approaches: firstly, it provides a dataset that can be used for evaluation and training;
secondly, we conjecture that the quality of artificial intelligence approaches will be massively improved if they are applied on top of a large set of guaranteed-perfect alignments.
This is based on two observations:
\begin{compactitem}
 \item The more alignments we know, the easier it is to find new ones.
Consider a typical formalization in system $S_1$ that introduces a new concept $c$ relative to some known ones.
If perfect alignments to system $S_2$ for the known concepts have already been established, it becomes relatively easy to find the formalization of $c$ in $S_2$, and add an alignment for it.
\item It is very difficult to get alignment-finding off-the-ground.
Because the foundations of $S_1$ and $S_2$ are often very different, almost nothing looks particularly similar in the absence of any alignments.
Deeper alignments will become more apparent only when alignments for fundamental concepts such as booleans, sets and numbers are established.
\end{compactitem}

\paragraph{Leveraging Alignments}
Here we focus on using alignment to translate expressions between libraries.
However, alignments also allow for a variety of other services such as simultaneous browsing and searching of multiple corpora, aiding premise selection, statistical analogies or refactoring of formal corpora; for more details we refer to~\cite{MueGauKal:cacfms17}.

\paragraph{Overview}
In Sect.~\ref{sec:mmt}, we recap the relevant preliminaries about alignments and the \mmt framework.
Then we describe the interface theories and our alignments in Sect.~\ref{sec:interfaces} and~\ref{sec:alignments}, respectively.
It is worth noting that these results were not found in that order: we first collected a large set of alignments between the formal libraries, then we constructed interface theories by abstracting over these alignments.
In Sect.~\ref{sec:implementations} we apply our design to concrete expression translations.
We conclude in Sect.~\ref{sec:conclusion} with a vision of a future collaborative effort expanding our work with many more interface theories and alignments.
 
 \section{Preliminaries}\label{sec:mmt}
 \subsection{Alignments}

\paragraph{} We speak of \emph{alignment} if the same (or a closely related) concept occurs in different libraries, possibly with slightly different names, notations, or formal definitions. Two aligned symbols thus can be thought to represent ``the same" abstract mathematical concept while differing only in implementation details. The notion of alignments is intentionally broad as to cover a wide variety of ways in which two symbols can be seen as ``morally the same".

For example, consider untyped and typed equality, e.g. in a set-theoretical and a dependently typed language respectively:
\begin{align*}
\textsf{eq}_1 &: \textsf{Set}\to\textsf{Set}\to\textsf{bool} \\
\textsf{eq}_2 &:(T:\textsf{Type})\to T\to T\to \textsf{bool}.
\end{align*}

Obviously, these two symbols are not easily interchangeable since they don't even have the same type; nevertheless, if we assume either symbol belonging to a distinct formal system, they clearly represent the same mathematical concept (namely \emph{equality}). Furthermore, if we want to translate any expression from one system to the other, we will clearly need to replace occurrences of one kind of equality by the respective other one.

It thus makes sense to think of both equalities as being \emph{aligned}.

For example, the alignment between $\textsf{eq}_1$ and $\textsf{eq}_2$ can be given as
	\[\textsf{system1:eq}_1\textsf{ system2:eq}_2\textsf{ arguments=``(1,2)(2,3)" direction=``both"}\]
For details on alignments we refer to~\cite{MueGauKal:cacfms17}, where alignments are classified, discussed and their implementations in \mmt are described in detail.

\subsection{The \mmt Framework}

\mmt \cite{RabKoh:WSMSML13,rabe:howto:14} is a wide-coverage representation language for formal mathematical knowledge.
It can be seen as a triple of the fragment of OMDoc \cite{Kohlhase:OMDoc1.2} dealing with logical and related knowledge, a rigorous semantics for that fragment, and a mature implementation.
\ommt is designed to
\begin{compactitem}
 \item avoid a commitment to any particular foundational type system or logic,
 \item allow for highly modular representations of foundational systems or domain knowledge,
 \item support interoperability across foundations, tools, and libraries.
\end{compactitem}
That makes it an ideal choice for describing interface theories.

\begin{figure}[ht]\centering
  \fbox{\includegraphics[width=0.45\textwidth]{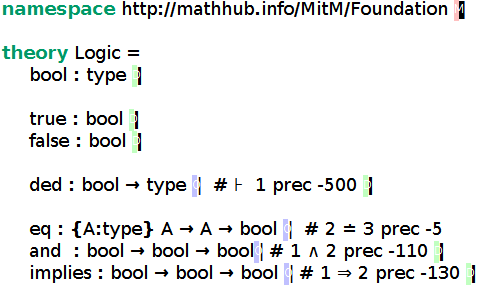}}
  \caption{An interface theory for logic in \mmt}\label{fig:mmtlogic}
\end{figure}

In the sequel, we explain by example the key features of \mmt that are relevant for our purposes.
Figure \ref{fig:mmtlogic} shows an example theory in \mmt surface syntax that can serve as interface for basic logical constants.

Here, a new theory \expr{Logic} is declared with URI \textsf{http://mathhub.info/MitM/Foundation?Logic}, which is composed of the namespace declared in the first line and the name of the theory. Afterwards, three constants are declared and given a type; namely the type of booleans \expr{bool} of type \expr{type}. The symbol \expr{type} is provided via a logical framework. The constants \expr{true} and \expr{false} are declared to be of type \expr{bool}.

In general, \mmt constants have the form \expr{c [:TYPE] [=DEF] [\#NOT]}, though we will not need definitions in this paper. The individual components (type, definition, notation, respectively) are all optional and can be provided in any order.

The symbol \expr{ded} will serve as a function from propositions to the type of their proofs to make use of the Curry-Howard correspondence. It is given the appropriate type \expr{bool \to type} and a notation \expr{\vdash\textsf{ 1 }}. The latter allows for writing \expr{\vdash\textsf A} for the type of proofs of a proposition $A$. Furthermore, the notation is provided with a precedence, to allow for omitting brackets.

The next symbols provide a typed equality and the basic logical connectives. The curly braces denote the dependent function type, providing for example \expr{eq} with the type $\prod_{\textsf{A:type}}\textsf A\to\textsf A\to\textsf{bool}$. The notation \expr{2 \doteq 3} omits the first argument \expr{A:type}, leaving it implicit and to be inferred by the system.

\paragraph{}
We can use this theory as one (out of many possible) meta-theory for various interface theories.
\begin{figure}[ht]\centering
  \fbox{\includegraphics[width=0.5\textwidth]{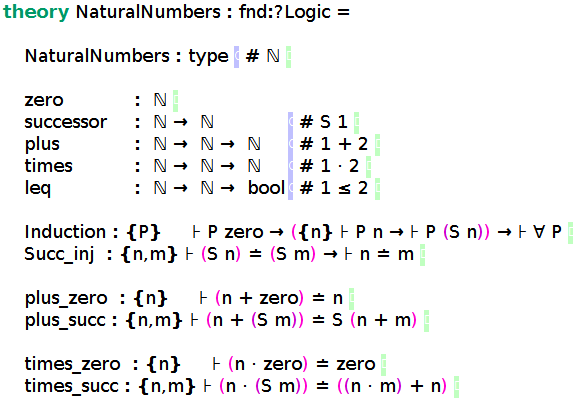}}
  \caption{An (excerpt of an) interface theory for natural numbers with \textsf{Logic} as meta-theory}\label{fig:natinterface}
\end{figure}
Figure \ref{fig:natinterface} shows an example of a simple theory of natural numbers, using the theory \expr{Logic} and the symbols declared therein.

\paragraph{} Using this modular approach as well as the foundation-independent nature of \ommt, the core primitives of various formal systems (such as interactive theorem provers)
can be (and in some cases have been, see e.g. \cite{OAF:on} or \cite{KohRab:qrtpflmk15} for the big picture) represented in \ommt alongside their libraries by using a formalization of the former as 
meta-theory for the theories in the latter, making either equally accessible to the system. This is what makes \mmt a ``meta-system'' suitable for our purposes, instead of the $n+1$st competing standard.



 \section{Interface Theories}\label{sec:interfaces}
 While alignments have the big advantage that they are cheap to find and implement, they have the disadvantage of not being very expressive; in fact, the definition of alignment itself is (somewhat intentionally) vague\footnote{In \cite{MueGauKal:cacfms17}, we attempt to classify alignments in more detail to contribute to a solution.}. In particular when a translation needs to consider foundational aspects beyond the individual system dialects, alignments alone are insufficient. This is where \emph{interface theories} (\cite{KohRabSac:fvip11,CarFarKoh:rsckmt14}) come into play: given different implementations of the same mathematical concept, their interface theory contains only those symbols that are a) common to all implementations and b) necessary to \emph{use} the concept (as opposed to formalizing it in detail)  -- which in practice turn out to be the same thing.

\subsection{Example: Natural Numbers}

In Mizar~\cite{mizar}, which is based on Tarski-Grothendieck set theory, the set of natural numbers \expr{NAT} is defined as \textsf{omega}, the set of finite ordinals. Arithmetic operations are defined directly on those.

In contrast, in PVS~\cite{OwRu92} natural numbers are defined as a specific subtype of the integers, which in turn are a subtype of the rationals etc. up to an abstract type \textsf{number} which serves as a maximal supertype to all number types. The arithmetic operations are inherited from a subtype \textsf{number\_field} of \textsf{number}.

These are two fundamentally different approaches to describe and implement an abstract mathematical concept, but for all practical purposes the concept they describe is the same; namely the natural numbers. The interface theory for both variants would thus only contain the symbols that are relevant to the abstract concept itself, independent of their specific implementation -- hence, things like the type of naturals, the arithmetic operations and the Peano axioms. The interface theory thus provides everything we need to work with natural numbers, and at the same time everything we know about them independently of the logical foundation or their specific implementation within any given formal system.

\paragraph{}However, there is an additional layer of abstraction here, namely that in stating that the natural numbers in Mizar are the finite ordinals we have already ignored the system dialect (in the sense of p.\ref{lbl:dialect}). This step of abstraction (from the concrete definition using only Mizar-specific symbols) yields another interface theory for finite ordinals, which in turn can be aligned not just with Mizar natural numbers, but also e.g. with MetaMath~\cite{MetaMath:on}, which is built on ZFC set theory.

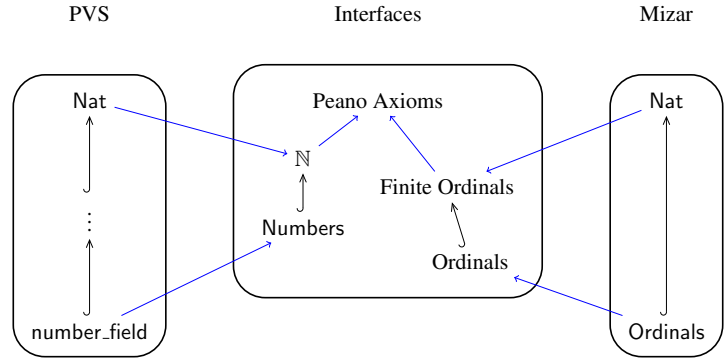
\begin{figure}
\begin{center}
\resizebox{0.6\textwidth}{!}{
\begin{tikzpicture}

\node (PVS) at (1,8.5) {PVS};

\node (PVNat) at (1,7) {\textsf{Nat}};
\node (PVdots) at (1,5) {$\vdots$};
\node (PVnf) at (1,3) {\textsf{number\_field}};
\draw[\arrowtipmono-\arrowtip] (PVnf) -- (PVdots);
\draw[\arrowtipmono-\arrowtip] (PVdots) -- (PVNat);
 \node[draw,thick,fit=(PVNat) (PVdots) (PVnf),
                   rounded corners=.55cm, inner sep=5pt] (PVScloud) {};

\node (Int) at (6,8.5) {Interfaces};

\node (ST) at (7.6,4.2) {Ordinals};

\node (Reals) at (4.7,4.8) {\textsf{Numbers}};
\node (Nat) at (4.7,6) {$\mathbb N$};
\draw[\arrowtipmono-\arrowtip] (Reals) -- (Nat);

\node (FOrd) at (7.2,5.5) {Finite Ordinals};

\node (PA) at (6,7) {Peano Axioms};
\draw[->,blue] (FOrd) -- (PA);
\draw[->,blue] (Nat) -- (PA);
\draw[\arrowtipmono-\arrowtip] (ST) -- (FOrd);

 \node[draw,thick,fit=(ST) (Reals) (Nat) (FOrd) (PA),
                   rounded corners=.55cm, inner sep=10pt] (PVScloud) {};

\node (Miz) at (11,8.5) {Mizar};

\node (MNat) at (11,7) {\textsf{Nat}};
\node (MOrd) at (11,3) {\textsf{Ordinals}};
\draw[\arrowtipmono-\arrowtip] (MOrd) -- (MNat);

 \node[draw,thick,fit=(MNat) (MOrd),
                   rounded corners=.55cm, inner sep=5pt] (PVScloud) {};

\draw[->,blue] (MOrd) -- (ST);
\draw[->,blue] (PVNat) -- (Nat);
\draw[->,blue] (PVnf) -- (Reals);
\draw[->,blue] (MNat) -- (FOrd);

\end{tikzpicture}
}
\end{center}
\caption{A graph showing different theories for natural numbers}\label{fig:intgraph}
\end{figure}

Figure \ref{fig:intgraph} illustrates this situation. Blue arrows point from more detailed theories to their interfaces. The arrows from PVS or Mizar to interfaces merely strip away the system dialects; the arrows within Interfaces abstract away more fundamental differences in definition or implementation.

\paragraph{} Consider again Figure \ref{fig:natinterface}, a possible interface theory for natural numbers. Note, that symbols such as \textsf{leq} \emph{could} be defined, but don't actually need to be. Since they are only interfaces, all we need is for the symbols to exist. 
 
 In fact, the more abstract the interface, the less we \emph{want} to define the symbols -- given that there's usually more than one way to define symbols, definitions are just one more thing we might want to abstract away from completely.
 
 The symbols in this interface theory can then be aligned either with symbols in other formal systems directly, or with additional interfaces in between, such as a theory for Peano arithmetic, or the intersection of all inductive subsets of the real numbers, or finite ordinals or any other possible formalization of the natural numbers.

\subsection{Additional Interface Theories}

The foundation independent nature of \mmt allows us to implement interface theories with almost arbitrary levels of detail and for vastly different foundational settings.

\begin{figure}[ht]\centering
  \fbox{\includegraphics[width=0.6\textwidth]{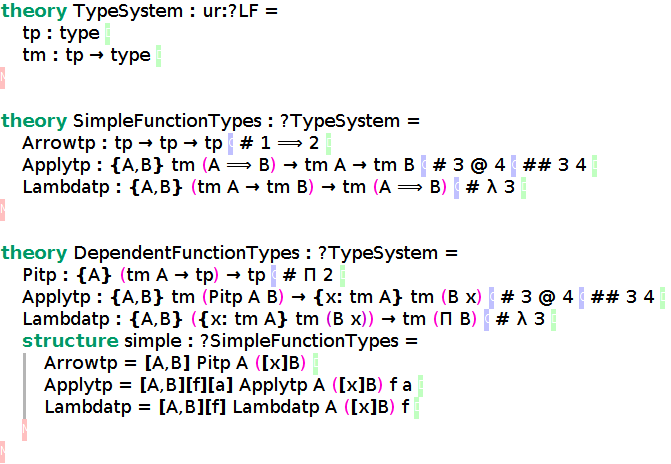}}
  \caption{Interface theories for type-theoretical foundations\label{fig:typeinterface}}
\end{figure}

We have started a repository of interface theories specifically for translation purposes~\cite{MitMInter:on} and also aligned to already existing interfaces (as in the case of arithmetics, see below) in a second and third \mathhub repository~\cite{MitMsmglom:on} and~\cite{MitMFoundation:on} extending them when necessary. 
Crucially, this interface repository contains interface theories for basic type-related symbols like the function type constructors (see Figure \ref{fig:typeinterface}\footnote{This interface theory, like most formalizations of foundations, uses types as terms (via the symbols \expr{tp} and \expr{tm}), whereas the interface theories above (like \expr{NaturalNumbers}) use the universe of types provided by the logical framework directly. During translation, special \emph{higher-order abstract syntax rules} take care of eliminating or inserting the corresponding symbols appropriately to make aligning between the two formalization levels possible.}), that are aligned with the respective symbols in HOL Light and PVS. These symbols are so basic as to be primitive in systems based on type theory, and consequently they occur in the vast majority of expressions. To have these symbols aligned is strictly necessary to get any further use of alignments off the ground.

Here, a \emph{structure} is used to include the theory for simple function types in the theory for dependent function types, while providing definitions for the symbols in terms of the latter. This automatically yields a translation from the simple to the dependent variant. 

Table~\ref{tbl:alignmenttotal} shows the total number of alignments we found for PVS, HOL Light and Mizar. The following are some additional examples of mathematical areas covered by the current interface theories:

 \begin{itemize}
  \item[\textbf{Calculus}] contains the following 3 subinterfaces:
  \begin{itemize}
  	\item[\textbf{Limits}] currently contains 17 symbols related to sequences and limits, including \expr{metric\_space} and \expr{complete} (metric spaces and their completeness). 
  	\item[\textbf{Differentiation}] currently contains 4 symbols, namely differentiability in a point and on a set and the derivative in a point and as a function
  	\item[\textbf{Integration}] currently contains 6 symbols, namely integrability and the integral over a set for Riemann, Lebesgue and Gauge-integration.
  \end{itemize}
  \item[\textbf{Arithmetics}] is an already existing interface theory from ~\cite{MitMsmglom:on}. It contains the interfaces for below number arithmetics (each split into two interfaces for the basic number type definitions and the arithmetics on them). 
  \begin{itemize}
    \item[\textbf{Complex Numbers}] currently contains 11 symbols for complex numbers aligned to their counterparts in HOL Light, PVS and Mizar. Besides the usual arithmetic operations similar to \expr{NaturalNumbers}, it contains \expr{i} (the imaginary unit), \expr{abs} (the modulus of a complex number) and \expr{Re}, \expr{Im}  (the real and imaginary parts of a complex number). 
    \item[\textbf{Integers}] currently contains 9 symbols for the usual arithmetic operations on integers and for comparison between two integers. 
    \item[\textbf{Natural Numbers}] currently contains 21 symbols and is already described above. 
    \item[\textbf{Real Numbers}] currently contains 15 symbols, again very similar in nature to the other number spaces. 
  \end{itemize}
  \item[\textbf{Lists}] currently contains the 13 most important symbols for lists, including \expr{head}, \expr{tail}, \expr{concat}, \expr{length} and \expr{filter} (filter a list using another list) as well as some auxiliary definitions. 
  There are no lists in Mizar, instead finite sequences are used. These however deserve their own interface.
  \item[\textbf{Logic}] is an already existing interface in the~\cite{MitMFoundation:on} repository. It contains 9 symbols for boolean algebra that are all perfectly aligned to HOL Light, PVS and Mizar the like and sometimes also to Coq.
  \item[\textbf{Sets}] is again an already existing interface theory from ~\cite{MitMsmglom:on} split into many subtheories. Currently, 28 of the contained symbols have been aligned. sets contains symbols for typed sets as in a type theoretical setting, including axioms and theorems. Here we have the most alignments so far. It also contains the following two interfaces:
  \begin{itemize}
  	\item[\textbf{Relations}] currently contains 23 symbols for alignments to relations and their properties, including orders. 
  	\item[\textbf{Functions}] currently contains 7 symbols for alignments to functions and their relations, that are not already contained in relations. 
  \end{itemize}
  \item[\textbf{Topology}] currently contains 25 symbols for both general topological spaces as well as the standard topology on $\mathbb R^n$ specifically. Since this yields additional difficulties, it will be examined in more detail in the next     section.
 \end{itemize}
 \paragraph{} As additional examples, the interface theories for limits and sets can be found in Appendix \ref{sec:interface_examples}, Figure \ref{fig:anatheory} and Figure \ref{fig:setstheory}.

 
 \section{Alignments}\label{sec:alignments}
 \subsection{Sample Alignments}
We manually combed through libraries of HOL Light, PVS, Mizar and Coq to find alignments.
Specifically, we picked the mathematical areas of numbers, sets (as well as lists), abstract algebra, calculus, combinatorics, logic, topology, and graphs as a sample.
This produced around 900 declarations overall, from which we constructed the interface theories presented in Sect. \ref{sec:interfaces}.

\begin{table}[ht]\centering\tiny
	\begin{tabular}{|c|c|c|c|c|c|}\hline
		Interface & PVS (Standard)& HOL Light (Standard)& Mizar (Standard) & Coq (Standard)\\\hline
		\textsf{nat\_lit} & \textsf{naturalnumbers?naturalnumber} & \textsf{nums?nums} & \textsf{ORDINAL1?modenot.6}& \textsf{Coq.Init.Datatypes?nat}\\
		\textsf{succ} & \textsf{naturalnumbers?succ} & \textsf{nums?SUC} & \textsf{ORDINAL1?func.1} & \textsf{Coq.Init.Nat?succ}\\
	 	\textsf{addition} & \textsf{number\_fields?+} & \textsf{arith?ADD} & \textsf{ORDINAL2?func.10} & \textsf{Coq.Init.Nat?add}\\
		\textsf{multiplication} & \textsf{number\_fields?*} & \textsf{arith?MULT} & \textsf{ORDINAL2?func.11}& \textsf{Coq.Init.Nat?mul} \\
		\textsf{lethan} & \textsf{number\_fields?$<=$} & \textsf{arith?$<=$} & \textsf{XXREAL\_0?pred.1}& \textsf{Coq.Init.Nat?leb} \\\hline
	\end{tabular}
	\caption{Alignments to the interface theory \textsf{NaturalNumbers} (libraries in brackets)}\label{tbl:natalignments}
\end{table}

\begin{table}[ht]\centering\tiny
	\begin{tabular}{|c|c|c|c|c|}\hline
		Interface & PVS (NASA\footnotemark ) & HOL Light (Standard)& Mizar (Standard) &Coq (coq-topology\footnotemark)\\\hline
		\textsf{topology} & \textsf{topology\_prelim?topology} & \textsf{topology?topology} & \textsf{PRE\_TOPC?modenot.1}& \textsf{TopologicalSpaces?TopologicalSpace} \\
		\textsf{open} & \textsf{topology?open?} & \textsf{topology?open\_in} & \textsf{PRE\_TOPC?attr.3} & \textsf{TopologicalSpaces?open} \\
		\textsf{closed} & \textsf{topology?closed?} & \textsf{topology?closed\_in} & \textsf{PRE\_TOPC?attr.4} & \textsf{TopologicalSpaces?closed}\\
		\textsf{interior} & \textsf{topology?interior} & \textsf{topology?interior} & \textsf{TOPS\_1?func.1}& \textsf{InteriorsClosures?interior}   \\
		\textsf{closure} & \textsf{topology?Cl} & \textsf{topology?closure} & \textsf{PRE\_TOPC?func.2 }& \textsf{InteriorsClosures?closure}   \\\hline
	\end{tabular}
	\caption{Alignments to the interface theory \textsf{Topology} (libraries in brackets)}\label{tbl:topalignments}
\end{table}

For example, Table \ref{tbl:natalignments} and~\ref{tbl:topalignments} show some of these alignments for the interface theories for natural numbers from Figure \ref{fig:intgraph} and topology. The URIs for PVS consist of the name of the containing theory in PVS (the alignments from the first table are to PVS's prelude, the alignments from the second are to the NASA library) and the name of the symbol within the interface, separated by a question mark\footnotemark. The ? appearing at the end of some of the URIs are actually a part of the name of the symbol in PVS. The URIs for HOL Light (prelude) and Mizar (mll) consist of the filename and the symbol name within the file.  

 \addtocounter{footnote}{-3}
 \stepcounter{footnote}\footnotetext{See http://github.com/nasa/pvslib}\label{fn:nasa}
 \stepcounter{footnote}\footnotetext{See http://github.com/coq-contribs/topology}\label{fn:coq-topology}
\stepcounter{footnote}\footnotetext{The general structure of MMT URIs is \textsf{$<$Namespace$>$?$<$Theory name$>$?$<$Symbol name$>$}, see e.g. \cite{RabKoh:WSMSML13}}
As mentioned earlier, alignments in topology pose some additional difficulties. Firstly, HOL Light defines a topology on some \emph{subset} of the universal set of the type, whereas PVS defines it on the universal set of the type directly. Thus, the alignment from HOL Light to the interface theory is unidirectional. Secondly, Mizar does not define the notion of a topology, but instead the notion of a topological space. Therefore, we align them to two different symbols in the interface (\expr{topology} and \expr{topological\_space}) and define \expr{topological\_space} based on \expr{topology}, so that \mmt can still translate expressions based of these two definitions. 

The set of all alignments we found can be inspected at \url{https://gl.mathhub.info/alignments/Public/tree/master/manual}. 
\subsection{Alignment Directions}
For translation purposes, we distinguish among three kinds of alignments (which are different from the categories in \cite{MueGauKal:cacfms17}) based on the possible directions of translations between the concepts in the interface theory and the prover library.
\paragraph{Bidirectional Alignments} This includes perfect alignments. For example, the translation between the definitions of set union in the interface theory and PVS library is bidirectional:
\subparagraph{Interface}\expr{union\ :\ \{A\}\ set\ A \to set\ A \to set\ A\ \#\ 2\cup \ 3}
\subparagraph{PVS}\expr{union(a,\ b):\ set\ =\ {x\ |\ member(x,\ a)\ OR\ member(x,\ b)}}
\paragraph{Unidirectional Alignments} This includes alignments up to totality of functions, alignments up to certain arguments and alignments up to associativity. For example, in PVS the operator \expr{+} is used universally for all number fields ($\mathbb N$, $\mathbb R$, $\mathbb C$), but in the interface theory \expr{plus} is defined for each type. Thus we can only translate from the interface theory to PVS.
\paragraph{Other Alignments} Due to the limitation of \mmt's current implementations, there are some alignments which cannot be used at the moment but are potentially directional. For example, the \expr{>} operator in Mizar is not explicitly declared, but instead defined as the so-called \expr{antonym} of the \expr{<=} operator. It is redirected to the \expr{<=} operator whenever used: \[\expr{antonym\ a\ >\ b\ for\ a\ <=\ b;}\] However, since \mmt cannot handle alignments up to negation, this cannot be used for translation yet.

\begin{table}[ht]\centering\footnotesize
	\begin{tabular}{|c|c|c|c|c|}
	  \hline 
	   Topic & HOL Light & PVS & Mizar & Coq\\
		\hline \textsf{Algebra}& \textsf{0/0} & \textsf{18/1} & \textsf{17/0} & \textsf{14/0}\\
		\hline \textsf{Calculus}& \textsf{15/0} & \textsf{14/0} & \textsf{16/0} & \textsf{5/15}  \\
		\hline \textsf{Categories} & \textsf{0/0} & \textsf{0/0} & \textsf{9/1} & \textsf{5/0}\\
		\hline \textsf{Combinatorics}& \textsf{24/0} & \textsf{15/0} & \textsf{1/0} & \textsf{1/0}  \\
		\hline \textsf{Complex Numbers} & \textsf{9/2} & \textsf{4/6} & \textsf{7/2} & \textsf{11/2}\\
		\hline \textsf{Graphs} & \textsf{5/5} & \textsf{17/0} & \textsf{20/0} & \textsf{7/2}\\
		\hline \textsf{Integers}& \textsf{10/0 }& \textsf{0/0 }& \textsf{5/2} & \textsf{47/3} \\
		\hline \textsf{Lists}& \textsf{16/0} & \textsf{9/0} & \textsf{8/0} & \textsf{36/2} \\
		\hline \textsf{Logic}& \textsf{7/0} & \textsf{7/5} & \textsf{7/0}& \textsf{24/1} \\
		\hline \textsf{Natural Numbers} & \textsf{19/0} & \textsf{8/10} & \textsf{9/0}& \textsf{34/1}  \\
		\hline \textsf{Polynomials} & \textsf{4/0} & \textsf{1/0} & \textsf{7/0}& \textsf{0/0}  \\
		\hline \textsf{Rational Numbers} & \textsf{0/14} & \textsf{2/11} & \textsf{0/10}& \textsf{14/3}  \\
		\hline \textsf{Real Numbers}  & \textsf{13/2} & \textsf{3/10} & \textsf{7/4} & \textsf{12/2}\\
		\hline \textsf{Relations}  & \textsf{4/0} & \textsf{16/5} & \textsf{18/3} & \textsf{1/12}\\
		\hline \textsf{Sets}  & \textsf{23/0} & \textsf{28/0} & \textsf{18/0} & \textsf{19/0}\\
		\hline \textsf{Topology} & \textsf{15/0} & \textsf{10/0} & \textsf{9/0}& \textsf{17/1} \\
		\hline \textsf{Vectors} & \textsf{13/0} & \textsf{7/0} & \textsf{15/0}& \textsf{0/0} \\
		\hline \textbf{Sum}  & \textbf{177/23} & \textbf{159/48} & \textbf{173/22}& \textbf{240/42} \\ 
		\hline 
	\end{tabular}
	\caption{Number of bidirectional/unidirectional alignments per library}\label{tbl:alignmenttotal}
\end{table}

\subsection{Causes for Imperfect Alignments}
The most common causes for imperfect alignments are subtyping and partiality.
\paragraph{Subtyping} In the PVS library, the arithmetic operations on all the number fields are defined on a common supertype \expr{numfields}. Therefore, a translation from PVS to the other two languages may not be viable. 
\paragraph{Partiality} The result of division by zero in the libraries of HOL Light and Mizar is defined as zero; in PVS, however, the divisor must be nonzero. Therefore, certain theorems in HOL Light and Mizar involving division no longer hold in PVS, and the translation is unidirectional. 

\paragraph{}
In order to translate an expression from one library to another, the concepts in the expression must at least exist in both libraries. This creates the need to inspect the intersection of the concepts in these libraries.
Table~\ref{tbl:intersectionConcepts} gives an overview of the library intersection for various interface theories.
\begin{table}[ht]\centering\footnotesize	
	\begin{tabular}{|c|c|c|c|c|}
		\hline Topic & 1 System & 2 Systems & 3 Systems & 4 Systems\\
		\hline \textsf{Algebra}& \textsf{17} & \textsf{9} & \textsf{5} & \textsf{0}\\
		\hline \textsf{Calculus}& \textsf{35} & \textsf{7} & \textsf{8} & \textsf{0}  \\
		\hline \textsf{Categories} & \textsf{4} & \textsf{5} & \textsf{0} & \textsf{0}\\
		\hline \textsf{Combinatorics}& \textsf{25} & \textsf{6} & \textsf{0} & \textsf{1}  \\
		\hline \textsf{Complex Numbers} & \textsf{10} & \textsf{5} & \textsf{3} & \textsf{3}\\
		\hline \textsf{Graphs} & \textsf{72} & \textsf{6} & \textsf{3} & \textsf{0}\\
		\hline \textsf{Integers}& \textsf{52 }& \textsf{2 }& \textsf{7} & \textsf{0} \\
		\hline \textsf{Lists}& \textsf{28} & \textsf{8} & \textsf{9} & \textsf{0} \\
		\hline \textsf{Logic}& \textsf{18} & \textsf{0} & \textsf{2}& \textsf{5} \\
		\hline \textsf{Natural Numbers} & \textsf{53} & \textsf{2} & \textsf{10}& \textsf{2}  \\
		\hline \textsf{Polynomials} & \textsf{12} & \textsf{0} & \textsf{0}& \textsf{0}  \\
		\hline \textsf{Rational Numbers} & \textsf{11} & \textsf{4} & \textsf{2}& \textsf{7}  \\
		\hline \textsf{Real Numbers}  & \textsf{9} & \textsf{3} & \textsf{5} & \textsf{5}\\
		\hline \textsf{Relations}  & \textsf{21} & \textsf{15} & \textsf{4} & \textsf{0}\\
		\hline \textsf{Sets}  & \textsf{56} & \textsf{10} & \textsf{9} & \textsf{10}\\
		\hline \textsf{Topology} & \textsf{62} & \textsf{2} & \textsf{8}& \textsf{0} \\
		\hline \textsf{Vectors} & \textsf{25} & \textsf{5} & \textsf{0}& \textsf{0} \\
		\hline \textbf{Sum}  & \textbf{510} & \textbf{91} & \textbf{79}& \textbf{33} \\ 
		\hline 
	\end{tabular}
	\caption{Number of concepts found in exactly one, two, three or four systems}
	\label{tbl:intersectionConcepts}
\end{table}

%
%
%

 \section{An Implementation of Alignment-based Translations}\label{sec:implementations}
 We can use alignments to translate expressions between different representations of the same mathematical concept. We can do so by simply substituting all occurrences of a symbol $a_1$ by the aligned symbol $a_2$ if translation is possible in this direction. For example, consider the expression \expr{member(a,A)} in PVS, stating that $a\in A$ for some set $A$ of type $T$. In \omdoc, this corresponds to the term (using pseudo-URIs)
\[\expr{PVS?apply(PVS/sets?member, T, a, A).}\]

Since \expr{member} is aligned with \expr{IN} in HOL Light, we can simply substitute one by the other. Furthermore, the symbol \expr{PVS?apply} representing function application in PVS can be assumed to be aligned with HOL Light's function application, yielding the term
\[\expr{HOLLight?apply(HOLLight/Sets?IN, T, a, A)},\] which corresponds to the HOL Light expression \expr{a\, IN\, A} in \omdoc.

The additional components of alignments allow for stating additional translation instructions such as switching, adding or omitting arguments. 

\paragraph{} For larger differences in implementation for which alignments are insufficient, we instead use interface theories for the (more or less) specific implementation variants that abstract away the system dialects. This allows us to implement arbitrary elaborate translation mechanisms on a generic level, without needing to care about the system-specific details. 

Theory morphisms can connect and translate from more detailed implementations to more abstract ones, e.g. going from finite ordinals to a first-order theory of natural numbers. Other translation mechanisms such as one from a set-theoretical to a type-theoretical setting can be realized generically on interface theories and thus can be leveraged by multiple systems. The LATIN library \cite{CodHorKoh:palai11,LATIN:git} already provides a variety of logics, type theories, set theories and translations between them in \mmt that can be used for these purposes.

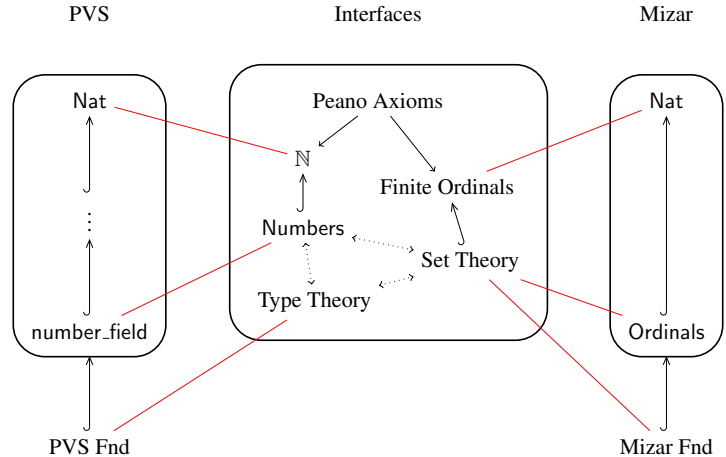
\begin{figure}
\begin{center}
\resizebox{0.6\textwidth}{!}{
\begin{tikzpicture}

\node (PVS) at (1,8.5) {PVS};
\node(PVSbot) at (1,2.7) {};

\node (PVSfnd) at (1,1) {PVS Fnd};
\draw[\arrowtipmono-\arrowtip] (PVSfnd) -- (PVSbot);

\node (PVNat) at (1,7) {\textsf{Nat}};
\node (PVdots) at (1,5) {$\vdots$};
\node (PVnf) at (1,3) {\textsf{number\_field}};
\draw[\arrowtipmono-\arrowtip] (PVnf) -- (PVdots);
\draw[\arrowtipmono-\arrowtip] (PVdots) -- (PVNat);

 \node[draw,thick,fit=(PVNat) (PVdots) (PVnf),
                   rounded corners=.55cm, inner sep=5pt] (PVScloud) {};

\node (Int) at (6,8.5) {Interfaces};

\node (TT) at (4.9,3.5) {Type Theory};
\node (ST) at (7.6,4.2) {Set Theory};
\draw[<->,dotted] (TT) -- (ST);

\node (Reals) at (4.7,4.8) {\textsf{Numbers}};
\node (Nat) at (4.7,6) {$\mathbb N$};
\draw[\arrowtipmono-\arrowtip] (Reals) -- (Nat);

\node (FOrd) at (7.2,5.5) {Finite Ordinals};

\node (PA) at (6,7) {Peano Axioms};
\draw[<-] (FOrd) -- (PA);
\draw[<-] (Nat) -- (PA);
\draw[\arrowtipmono-\arrowtip] (ST) -- (FOrd);

 \node[draw,thick,fit=(ST) (Reals) (Nat) (FOrd) (PA) (TT),
                   rounded corners=.55cm, inner sep=10pt] (PVScloud) {};

\node (Miz) at (11,8.5) {Mizar};

\node (Mizfnd) at (11,1) {Mizar Fnd};
\node(Mizbot) at (11,2.7) {};
\draw[\arrowtipmono-\arrowtip] (Mizfnd) -- (Mizbot);

\node (MNat) at (11,7) {\textsf{Nat}};
\node (MOrd) at (11,3) {\textsf{Ordinals}};
\draw[\arrowtipmono-\arrowtip] (MOrd) -- (MNat);

 \node[draw,thick,fit=(MNat) (MOrd),
                   rounded corners=.55cm, inner sep=5pt] (PVScloud) {};

\draw[red] (PVSfnd) -- (TT);
\draw[red] (Mizfnd) -- (ST);
\draw[red] (MOrd) -- (ST);
\draw[red] (PVNat) -- (Nat);
\draw[red] (PVnf) -- (Reals);
\draw[red] (MNat) -- (FOrd);

\draw[<->,dotted] (TT) -- (Reals);
\draw[<->,dotted] (ST) -- (Reals);

\end{tikzpicture}
}
\end{center}
\caption{A graph showing different translation paths between theories}\label{fig:graph}
\end{figure}

Figure \ref{fig:graph} shows possible relations on the theory level for the natural numbers example above: red lines stand for alignments, hooked arrows for theory inclusions, straight arrows for views and dotted arrows for other possible translations.

Given the various arrows, expression translation reduces to a simple graph search on the symbols (or, depending, on larger subexpressions) of the input.

\paragraph{} While we only translate single expressions instead of whole proofs, the former is a necessary first step for the latter. Furthermore, translating the expressions occuring in a proof yields a ``proof sketch'' in the target system, which -- we conjecture -- can often be enough to complete the proof there automatically.
 \paragraph{}We have implemented a prototypical expression translation in \mmt. It currently uses $\approx400$ alignments for translation purposes (out of $\approx1400$ in total) between HOLLight, PVS, Mizar and Coq (even though there is no \omdoc import from Coq as of yet). The translation mechanism will be exposed via the \mmt query language \cite{Rabe:qlfml12}\ednote{TODO!}. Besides alignments, it uses views, \mmt-structures and arbitrary other translations implemented in Scala and provided by \mmt plugins. Furthermore, it allows for grouping alignments so they are always used in conjunction, and prioritizing certain translations over others.

Crucially, it returns partial translations when no full translation to a target library can be found. These can be used for finding new alignments automatically by the techniques described in the introduction.

We furthermore have two libraries for interface theories; \textsf{MitM/interfaces} \cite{MitMInter:on} specifically for expression translation and, more generally \textsf{MitM} \cite{MitM:on}. The latter is additionally used in other projects for simlar purposes (\cite{KohKopMue:mmrdftg17} and \cite{ODKproposal:on}).

\begin{figure}[ht]\centering
  \fbox{\includegraphics[width=0.6\textwidth]{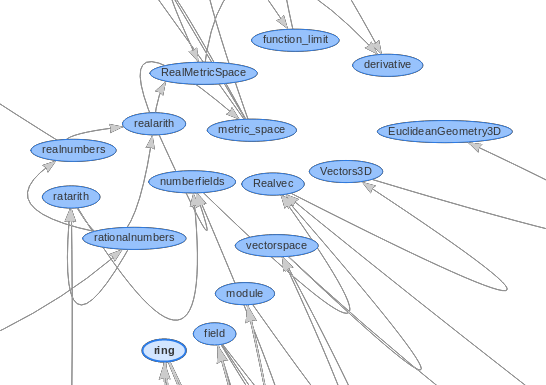}}
  \caption{A Small Part of the MitM Theory Graph}\label{fig:mitmgraph}
\end{figure}

Figure \ref{fig:mitmgraph} shows a small part of the theory graph of the \textsf{MitM} libraries. The full \textsf{MitM} theory graph can be explored on \mathhub\footnote{\url{https://mathhub.info/mh/mmt/graphs/tgview.html?type=archivegraph&graphdata=MitM}}.

The above can be (and has been) used to -- for example -- translate the expressions listed in Table \ref{tbl:translate}. The curly braces denote the context for the variables used. 

Note that even something like the application of a function entails a translation from function application in HOL Light to function application in PVS, since they belong to their respective system dialects. Furthermore, the translation from simple function types in HOLLight to the dependent $\Pi$-type in PVS is actually done on the level of interface theories, where simple function types are defined as a special case of dependent function types for this exact purpose. Unlike the latter two, the first example uses only primitive symbols in both systems that are not part of any external library. Meta-variables have been left out in the third example for brevity. 

Furthermore, an argument alignment was used in the third example, that switches the two (non-implicit) arguments -- namely the predicate and the list. Whereas this usually trips up automated translation processes or needs to be manually implemented, in our case this is as easy as adding (in this case) the key-value-pair \expr{arguments=``(2,3)(3,2)"} to the alignment.
\begin{table}[ht]\centering	\footnotesize
  \begin{tabular}{|c|c|}
 	\hline HOL Light & PVS \\ 
  	\hline \textsf{\{A:holtype, P:term A$\Longrightarrow$bool, a:term A\}$\vdash$P(a)} & \textsf{\{A:tp,P:expr $\Pi_{\textsf{a:A}}\textsf{boolean}$,a:expr A\}$\vdash$P(a)} \\ 
  	\hline \textsf{\{T:holtype,a:term T,A:term T$\Longrightarrow$bool\}a IN A} & \textsf{\{T:tp, a:expr T, A:expr T$\Longrightarrow$boolean\}member(a,A)} \\ 
  	\hline \textsf{FILTER (Abs x:bool. x) c :: b :: a :: NIL} & \textsf{filter (c :: b :: a :: null) ($\lambda$x : boolean. x)} \\ 
  	\hline 
  \end{tabular}
  \caption{Three Expressions Translated}\label{tbl:translate}
\end{table}

\section{Conclusion}\label{sec:conclusion}
We presented a case study that demonstrates that the approach of alignment-based translation of expressions between theorem prover libraries is feasible.
Concretely, we gave interface theories for five representative mathematical domains along several hundred alignments to three major libraries.

Our alignments are publicly available at \cite{Alignments-public:on}.
Our representations of libraries in MMT (including HOLLight, Mizar and PVS) can be found on \mathhub \cite{MathHub:on}.
We currently pursue additional imports from IMPS \cite{FaGu93}, TPS and Coq.

We see this paper as an initial, prototypical example of a community-wide effort to build a large library of interface theories and alignments.
We envision an interface library that subsumes all formal knowledge formalized in major theorem provers.
In the long run, new formalizations in any system should always be coupled with an interface theory that is uploaded to this central library.
(If successful, this agenda will eventually lead to an integration problem when multiple competing interface theories exist for the same domain.
But, while non-trivial, that problem is much simpler than the integration problem between libraries.)

Therefore, we call on the community to expand our set of interface theories and alignments, both in future publications and through pull requests to the above-mentioned git repository.
If an interface theory is already known or (as in our case) roughly known, finding more alignments is relatively easy.
For example, the alignments presented here were found by two strong undergraduate students in a total of around 230 hours with relatively little supervision.
Reporting the found alignments and parallelizing the work is easy using the standardized format presented in \cite{KalKohMue:alfms16}.
Therefore, we expect it will be easy to scale the manual approach up.
In fact, finding alignments can be a great exercise for students to familiarize themselves with a library.
%
\bibliographystyle{eptcsalpha}
\bibliography{biblio}


\begin{appendices}
 \section{More Examples of Interface Theories}\label{sec:interface_examples}
 \begin{figure}[ht]\centering
  \fbox{\includegraphics[width=0.9\textwidth]{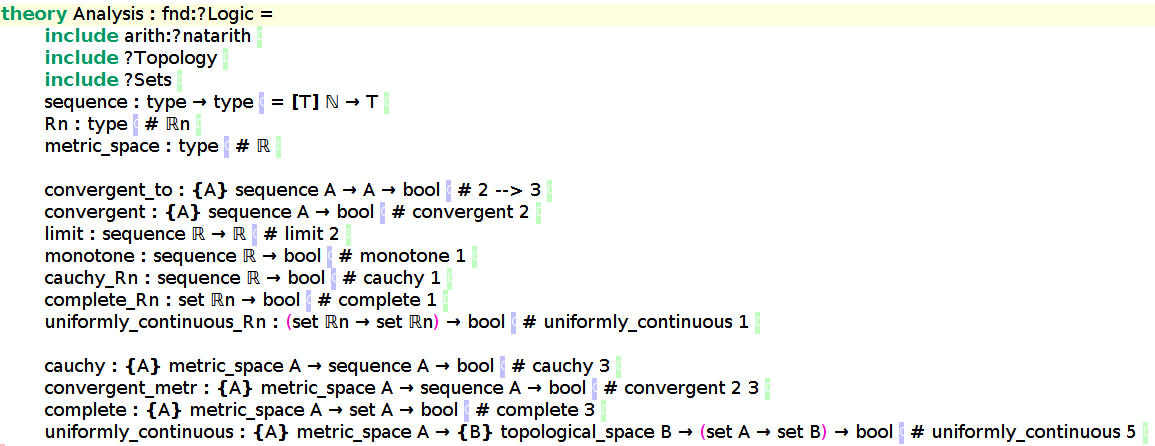}}
  \caption{An interface theory for analysis}\label{fig:anatheory}
\end{figure}

\begin{figure}[ht]\centering
  \fbox{\includegraphics[width=0.7\textwidth]{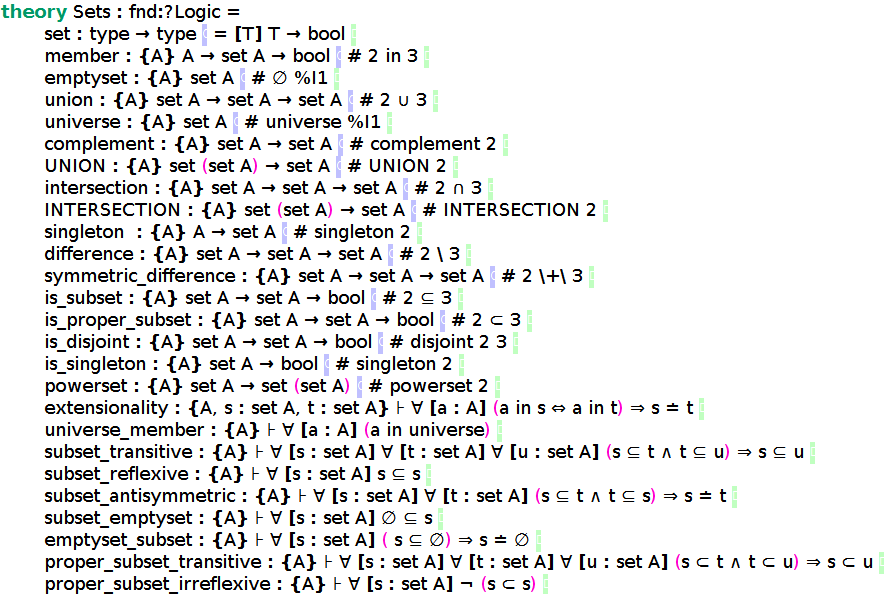}}
  \caption{An interface theory for typed sets}\label{fig:setstheory}
\end{figure}
\end{appendices}
\end{document}